\documentclass[10pt,a4paper,twocolumn]{article}

\usepackage[margin=0.85in]{geometry}
\usepackage{authblk}
\usepackage[T1]{fontenc}
\usepackage{lmodern}
\usepackage{graphicx}
\graphicspath{{./figures/}}
\usepackage{amsmath}
\usepackage{microtype}
\usepackage{amsfonts}
\usepackage[capitalize,noabbrev]{cleveref}
\usepackage{amssymb}
\usepackage[T1]{fontenc}
\usepackage[utf8]{inputenc}
\usepackage{cite}
\usepackage{subfig}
\usepackage{amsmath}
\usepackage{algorithm2e}
\usepackage{todonotes}
\RequirePackage{acronym}
\begin{document}
\title{A Hierarchical State-Machine-Based Framework for Platoon Manoeuvre Descriptions}
\author[1]{Corvin Deboeser}
\author[1,2]{Jordan Ivanchev}
\author[1]{Thomas Braud}
\author[2,4]{Alois Knoll}
\author[1,2]{David Eckhoff}
\author[3]{Alberto Sangiovanni-Vincentelli}
\affil[1]{TUMCREATE, Singapore}
\affil[2]{Technical University Munich,  Germany}
\affil[3]{University of California, Berkeley}
\affil[4]{Nanyang Technological University, Singapore}
%
\maketitle
\begin{abstract}
\boldmath{This paper introduces the SEAD framework that simplifies the process of designing and describing autonomous vehicle platooning manoeuvres. Although a large body of research has been formulating platooning manoeuvres, it is still challenging to design, describe, read, and understand them. This difficulty largely arises from missing formalisation. To fill this gap, we analysed existing ways of describing manoeuvres, derived the causes of difficulty, and designed a framework that simplifies the manoeuvre design process. Alongside, a Manoeuvre Design Language was developed to structurally describe manoeuvres in a machine-readable format. Unlike state-of-the-art manoeuvre descriptions that require one state machine for every participating vehicle, the SEAD framework allows describing any manoeuvre from the single perspective of the platoon leader. 
We hope that the SEAD framework will pave the way for further research in the area of new manoeuvre design and optimisation by largely simplifying and unifying platooning manoeuvre representation.}
\end{abstract}
\vspace{0.7em}
%
%
\vspace{0.55em}
\section{Introduction}
Multiple challenges must be solved along the way to a globally optimised and highly automated urban traffic system. 
While the past has seen a large body of research and technological innovation on AV, platooning as a concept is still in its early stages. With the capabilities of vehicular awareness and communication in place, research can now start building elaborate platooning strategies to leverage on those technological advancements.

A specialised body of research is concerned with formulation of collaborative driving manoeuvres whereas research is split into two major categories. 
The first category optimises driving behaviour on the vehicle's dynamics level for smooth and energy-efficient driving. 
The second category, in contrast, centres around collaboration and communication strategies, for instance, how an additional vehicle joins an existing platoon.
Within the second category, researchers have investigated specific cases of collaborative driving in platoons. 
Most papers focus on the specific details of one manoeuvre and optimise it to a great extent in terms of stability (the trait of a manoeuvre not to lead to dangerous traffic situations) or execution time. 

Although most papers draw on the description of manoeuvres through finite state machines, there is no consensus or convention among researchers how to represent manoeuvres. 
This heterogeneity aggravates the comparison of manoeuvres and requires researches to constantly adapt to new conventions. 
Besides, the description of manoeuvres through mere state machines requires multiple synchronised yet independent state machines, one for each participating vehicle. 
Designing and reading these state machines can become challenging even for simple manoeuvres. The objective of this work is to provide a framework that simplifies and formalises this description. 

The proposed framework follows four principles: Standardisation, Encapsulation, Abstraction, and Decoupling (SEAD). 
Standardisation ensures a common terminology among all areas within the framework and all researchers applying it. 
To take advantage of the repetitive occurrence of action-sequences in various contexts, the Encapsulation principle allows grouping such repetitive patterns into re-usable modules. 
Since some patterns may contain sub-patterns, the Abstraction principle leads to recursive encapsulation which allows considering manoeuvres and their building blocks on different levels of detail. 
The Decoupling principle, finally, separates the control of the manoeuvre from its execution. 
This allows describing manoeuvres exclusively from the control-perspective of the platoon leader while the framework ensures the correct execution behaviour. 
The SEAD framework may serve future research as a point of reference and tool to facilitate further manoeuvre investigations.

In summary, the contributions of this article are:
\begin{itemize}
\item We survey and structure the complex landscape of AV manoeuvre research, identify shortcomings, and derive requirements for a new framework.
\item We present SEAD, a novel AV manoeuvre framework to significantly simplify the process of manoeuvre modelling.
\item We make a library of manoeuvres publicly available, in both human-readable and machine-readable formats.
\end{itemize}

The remainder of this article is organised as follows:
First we give an extensive summary of related work in the domain of AV manoeuvre modeling (Section~\ref{sec:related}).
From this we derive in Section~\ref{sec:requirements} requirements and principles that a common framework should fulfil and follow.
Section~\ref{sec:sim} introduces our new SEAD framework.
Finally, in Section~\ref{sec:conclusion} we discuss possible extensions and conclude the article.

%
%

\section{Related Work}
\label{sec:related}

The concept of an Autonomous Highway Systems (AHS) was first introduced by Varaiya in 1993 under the name Intelligent Vehicle/Highway System (IVHS)~\cite{varaiya1993smart}, promising increases in safety and in highway capacity without the need of building new roads. 
These advantages emerge from two conceptual changes as compared to a conventional highway: 1) the application of vehicle platooning and 2) a global optimisation of traffic flow and travel times through a layered control structure.

A platoon is composed of a leader and at least one follower whereas, in most publications, the leader is the first vehicle in the upstream direction. 
The major impact of platooning for an increased highway capacity is the decreased inter-vehicle distance within platoons (intra-platoon distance d) as compared to the inter-vehicle distance outside of platoons (inter-platoon distance D)~\cite{varaiya1993smart}. As shown in~\cite{varaiya1993smart}, platooning may increase road capacity by up to a factor of three.

In the platooning logic, the leader sets the platoon speed and coordinates the manoeuvres performed by the platoon, for example splitting a platoon into two platoons (Split manoeuvre) or merging two platoons into one (Merge manoeuvre). The followers follow the preceding vehicle according to their Cooperative Adaptive Cruise Control (CACC) system and react upon commands issued by their platoon leader.

\begin{figure}[h]
	\centering

	\includegraphics[trim={0 0 0 0}, clip,width=0.4\textwidth]{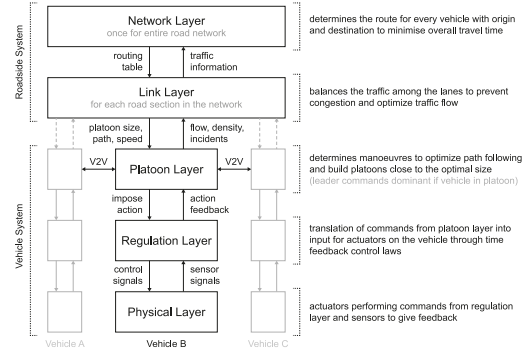}

	\caption{Control hierarchies of the platooning control system (summarized from~\cite{varaiya1993smart,godbole1995design,hsu1993protocol})}
	\label{fig:episode}

\end{figure}

Although platooning itself increases road capacity and vehicle throughput, it may become more effective when applying additional higher-order strategies that leverage on the vehicle’s communication capabilities. 
The interconnection of vehicles and infrastructure allows to control optimal platoon forming, driving, and splitting strategies and to optimise traffic globally instead of locally. 
The combination of such systems yields an AHS. 
Due to its complexity, the structure of the AHS is split into five hierarchical layers~\cite{varaiya1993smart,godbole1995design}. Each layer fulfils a mutually exclusive task as shown in \cref{fig:episode}.

At the highest level of the hierarchy, the Network Layer is responsible for optimizing the overall travel time of all vehicles and the traffic flow in the entire network~\cite{godbole1995design}. 
It is aware of all autonomous vehicles in the road network and the current and predictable traffic situation on every road~\cite{varaiya1993smart,godbole1995design}. 
To optimize travel time and traffic flow, it balances the traffic load on each road by determining the ideal path for each vehicle that travels from a defined origin to a defined destination~\cite{varaiya1993smart,ivanchev2016bisos}. 
For the case of IVHS, the network only consists of the highway.
The Link Layer is more decentralized and implemented by a controller for each road segment~\cite{varaiya1993smart}. 
The controller ensures a smooth traffic flow on its road segment by distributing the traffic vehicles among lanes~\cite{godbole1995design}. 
Besides determining a dedicated lane for each vehicle or platoon, it also determines the target size and velocity for platoons on that section~\cite{hsu1993protocol}. 
For an IVHS, the road segmentation is realized by dividing the highway into segments of equal length.
These two layers are implemented in the infrastructure and part of the roadside system. The layers below belong to the vehicle system, meaning that every vehicle is equipped with modules to realize the tasks of the three layers described below.

At the highest hierarchical level of the vehicle system is the Platoon Layer or Planning and Coordination-Layer. 
As a free vehicle, this layer determines actions to fulfil the path and lane directives imposed by the layers above~\cite{hsu1993protocol}. 
Part of this task is to determine lane changes, if a vehicle should join or leave a platoon~\cite{godbole1995design}. 
As a platoon leader, this layer coordinates the actions with vehicles that are associated with the platoon either as followers or potential joiners~\cite{varaiya1993smart}. 
As a platoon follower, the platoon layer collaboratively performs action protocols which are initiated by the platoon leader~\cite{varaiya1993smart,hsu1993protocol}.

The Regulation Layer and the Physical Layer are responsible for realising the trajectories computed by the hierarchically higher layers. 
Control loops in the Regulation Layer compute actionable commands for the actuators and minimise the errors reported by the sensors in the Physical Layer~\cite{hsu1993protocol}.

The Platooning Layer, which is the focus of this publication, contains collaborative driving logic: Platoons must be formed, maintained, and modified. 
Collaborative driving of multiple vehicles is described by manoeuvres. 
Manoeuvres encompass the driving and communication behaviour for all participating vehicles in the form of manoeuvre protocols. 
As shown in the subsequent sections, the description of manoeuvres is complex even for simple ones. 
Our work, thus, aims to increase the comprehensibility and facilitate the easy formulation of platoon manoeuvres.

\begin{figure}[h]
	\centering

	\includegraphics[trim={0 0 0 0}, clip,width=0.45\textwidth]{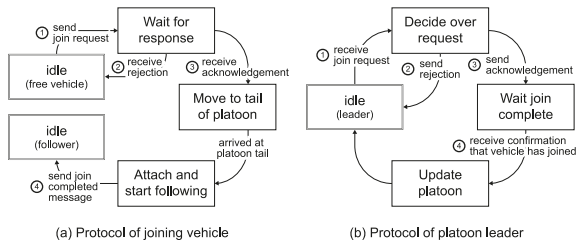}

	\caption{Example description of the JoinTail protocol as communicating FSM (simplification of protocol from~\cite{hsu1993protocol}). (a) behaviour of the joining vehicle, (b) behaviour of the platoon leader. Rectangles constitute the states and the arrows the transitions. States with double-stroke are Idle States (i.e. starting and ending states of a protocol) and the ones with a single stroke are action states (i.e. states where an action is performed). }
	\label{fig:jointail}

\end{figure}

A common way of describing these manoeuvre-protocols are communicating finite state machines e.g.~\cite{amoozadeh2015platoon, hsu1991design,rajamani2000demonstration}. 
In order to gain a better understanding of different manoeuvre description approaches we present a simplified example of the JOIN TAIL-manoeuvre from~\cite{hsu1993protocol} shown in \cref{fig:jointail}. 
The JOIN\ TAIL-protocol describes the process of a free vehicle joining an existing platoon at its tail through two CFSM.

When a vehicle (Vehicle A) decides to join a platoon, it sends a join request to the leader (Vehicle B) of the platoon (Transition 1, T1). B either rejects (T2) or acknowledges (T3) the request. 
In the case of rejection, both vehicles return to idle and the protocol terminates. 
If the request is acknowledged, A will wait for B to join the platoon. 
A moves to the tail of the platoon. 
Once arrived at the tail, A attaches to the platoon, starts following the preceding vehicle (i.e. switch into CACC mode), sends a message to B that the join is completed (T4), and transitions into the follower-Idle State. 
Upon receipt of the join-completion message, B updates the platoon information and returns to idle. 
The protocol execution is complete and terminates.

This description shows that state machines synchronise through messages that are sent between the vehicles. 
Since both vehicles operate through the same set of protocols, they can expect the other vehicle to behave synchronously.

The representation in \cref{fig:jointail} describes the behaviour as two separate FSM, each describing the protocol for one role within the manoeuvre. 
FSM (a) shows the transition of a vehicle from one Idle State into another. 
After extending the state machine by the transition from a free vehicle into a platoon leader through platoon formation, the entire behaviour could be explained in one role-agnostic FSM. 
This FSM operates on both vehicles independently and simultaneously. 
Both participating vehicles then operate following the same FSM schema, yet they are in different states during the manoeuvre and follow different paths through the FSM. 

\begin{figure}[h]
	\centering

	\includegraphics[trim={0 0 0 0}, clip,width=0.45\textwidth]{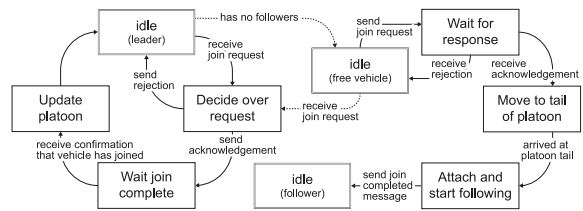}

	\caption{Combined FSM that represents the JoinTail-protocol as shown in Figure 9. The dashed arrows establish the connection between the two state machines.}
	\label{fig:jointailcombined}

\end{figure}

This approach was shown by~\cite{amoozadeh2015platoon} and is illustrated in \cref{fig:jointailcombined} as an extension to the example in \cref{fig:jointail}. 
The dashed arrows were added to combine the two FSM into one. 

In the same manner as in \cref{fig:jointailcombined}, it is possible to describe the entire behaviour of vehicles in one role- and manoeuvre-agnostic FSM, meaning that one FSM would describe the entire platooning behaviour of a vehicle for all roles.

Although FSM are the dominant way to describe manoeuvres in the current state of literature, researchers also applied other ways of illustration.
The authors of~\cite{hsu1993protocol} describe a manoeuvre in a role-agnostic process flow-chart. 
From this chart, they deduct role-specific state machines that are ultimately formulated in the COSPAN (coordination-specification-analysis) system. 
COSPAN formalises CFSM to prove certain mathematical state machine properties such as completeness or reachability~\cite{clarke1996computer}.

In~\cite{bengtsson2015interaction}, the logical flow of manoeuvres is also described as a role-agnostic flow-chart. 
Focusing on V2V communication, however, this paper also provides additional information flow-charts to describe the communication between the participating vehicles.
In addition to illustrations, the majority of papers describes the logical flow of events in manoeuvres through textual descriptions e.g.~\cite{amoozadeh2015platoon,hsu1993protocol,rajamani2000demonstration}.

As mentioned above, most publications describe manoeuvres as FSM whereas only a minority applies other visualization and description principles. 
However, although most manoeuvre descriptions use the same methodologies, there is still a vast heterogeneity among all existing publications. 
Two dominant sources of this heterogeneity are: 1) different levels of detail, and 2) verbally differing descriptions to represent identical actions.

\textbf{Different levels of detail:} Some manoeuvre descriptions show a level of detail that reaches the regulation layer with states such as Accelerate to Merge~\cite{hsu1993protocol} or Set Speed to 30 m/s~\cite{amoozadeh2015platoon}. 
Other descriptions, or sometimes even the same description, stay on a very high level within the platooning layer with states such as Car Splits~\cite{rajamani2000demonstration} or Move To Position~\cite{segata2014supporting}. 
Due to these differences of abstraction level, FSM that potentially describe the same behaviour are, in fact, significantly different. 

\textbf{Differing verbal descriptions:} Even if illustrations describe a manoeuvre on similar levels of detail, they may use different terminologies or different levels of verbal abstraction for the same action. 
For example, the Join Tail-protocol was described in~\cite{segata2014supporting} and in~\cite{hsu1993protocol} in similar logical ways. 
In both descriptions for the joining vehicle, a dedicated state equalises the speed of the platoon and the joining vehicle by decelerating or accelerating. 
While~\cite{segata2014supporting} describes this as Set Speed to 30 m/s and Catch-up and merge the platoon from the back,~\cite{hsu1993protocol} expresses the same process as Accelerate to Merge. This poses a challenge when formalising and comparing manoeuvres across different publications.

Besides these heterogeneities, we identified two additional factors that pose challenges when formulating manoeuvres. 
First, repetitiveness introduces pseudo-complexity, especially for those descriptions with high levels of conceptual detail. 
For instance, the process for a vehicle moving to a certain position requires both communication and physical action. 
The leader orders the joining vehicle to move to a defined position. 
When at that position, the joining vehicle informs the leader about the arrival. 
These action-communication-patterns may occur repetitively within a manoeuvre FSM. 

Second, since manoeuvres are mostly performed by two or more vehicles collaboratively, the behaviour for every participant is described in dedicated, role-dependent FSM. 
Reading these coupled state machines requires a substantial effort as the reader must manually synchronize the state machines to understand the collaborative process.

\begin{table*}[]
		\caption{Research publications regarding platooning manoeuvres}
		\label{Table:platooning_literature}
	\resizebox{\textwidth}{!}{%
	\begin{tabular}{|l|l|l|l|}
		\hline
		\textbf{Reference}                                                                                                                                            & \textbf{Manoeuvres}                                                                                                                               & \textbf{Additional Information}                                                                                                                                                        & \textbf{Cat*} \\ \hline
		\begin{tabular}[c]{@{}l@{}}Lu \& Hedrick 2003~\cite{lu2003longitudinal},\\   Lu et al. 2004~\cite{lu2004automated}\end{tabular}                      & Merge                                                                                                                                    & Design of controller for longitudinal merging, virtual platoon to pre-compute trajectory before manoeuvre                                                                     & P    \\ \hline
		Nowakowski et al. 2016~\cite{nowakowski2016operational}                                                                                              & Merge, Split                                                                                                                             & \begin{tabular}[c]{@{}l@{}}Platooning for non-autonomous trucks with CACC technology \\ \\ and V2V communication, manoeuvres performed by drivers\end{tabular}                & P    \\ \hline
		Segata et al. 2014~\cite{segata2014supporting}                                                                                                       & JoinMiddle (FSM)                                                                                                                         & \begin{tabular}[c]{@{}l@{}}Designing and investigating the JoinMiddle manoeuvre with \\ \\ various interference scenarios and communication packet loss\end{tabular}          & P    \\ \hline
		Hsu \& Liu 2004~\cite{hsu2004platoon}                                                                                                                & LaneChange                                                                                                                               & Proposes manoeuvre to LaneChangeAndFollow for higher efficiency                                                                                                               & P, R \\ \hline
		Hsu \& Liu 2008~\cite{hsu2008kinematic}                                                                                                              & LaneChange                                                                                                                               & Investigates LaneChange options: simultaneously and time delay with varying inter-platoon spacing                                                                             & P, R \\ \hline
		Kazerooni \& Ploeg 2015~\cite{kazerooni2015interaction}                                                                                              & Join, LaneReduction, GapOpen, GapClose                                                                                                   & Design of interaction protocols for LaneReduction and JoinMiddle, Analysis of velocity profiles                                                                               & P, R \\ \hline
		Rajamani et al. 2000~\cite{rajamani2000demonstration}                                                                                                & Join, Leave, LaneChange (FSM)                                                                                                            & Lateral and longitudinal control for Merge and Join, and LaneChange                                                                                                           & P, R \\ \hline
		Best 2012~\cite{best2012optimisation}                                                                                                                & -                                                                                                                                        & \begin{tabular}[c]{@{}l@{}}Rapid high-speed lane change for obstacle avoidance, \\ \\ proposal of open-loop controller for steering instead of emergency braking\end{tabular} & R    \\ \hline
		Choi \& Swaroop 2000~\cite{choi2001assessing}                                                                                                        & -                                                                                                                                        & Assessing coordinated emergency braking in platoons                                                                                                                           & R    \\ \hline
		Frankel et al. 1996~\cite{frankel1996safety}                                                                                                         & \begin{tabular}[c]{@{}l@{}}Merge, Split, \\   LaneChange\end{tabular}                                                                    & \begin{tabular}[c]{@{}l@{}}Proposal of safety-ensuring controllers for Merge, Split, \\ \\ and LaneChange as FSM, textual description of manoeuvres\end{tabular}              & R    \\ \hline
		Godbole et al. 1995~\cite{godbole1995design}                                                                                                         & Join, Leave (textual)                                                                                                                    & On- and off-ramp in AHS with dedicated lane for AVs                                                                                                                           & R    \\ \hline
		Goli \& Eskandarian 2014~\cite{goli2014evaluation}                                                                                                   & Merge, LaneChange                                                                                                                        & Multi-merge manoeuvre (multiple vehicles at one), lateral trajectory generation and execution via PID                                                                         & R    \\ \hline
		Huang \& Chen 2001~\cite{huang2001safe}                                                                                                              & Merge, Split                                                                                                                             & Investigates the safety of Merge and Split for emergency braking into different stages of the manoeuvres                                                                      & R    \\ \hline
		Kato et al. 2002~\cite{kato2002vehicle}                                                                                                              & \begin{tabular}[c]{@{}l@{}}Merge, Leave,\\   LaneReduction\end{tabular}                                                                  & Investigation of velocity profiles for merging, obstacle avoidance, leaving, and Stop\&Go                                                                                     & R    \\ \hline
		Li et al. 1997~\cite{li1997ahs}                                                                                                                      & -                                                                                                                                        & \begin{tabular}[c]{@{}l@{}}Longitudinal control laws with focus on safety that can be \\ \\ adjusted through parameters for changing external conditions\end{tabular}         & R    \\ \hline
		\begin{tabular}[c]{@{}l@{}}Murthy \& Masrur 2016~\cite{murthy2016braking}, \\ \\   Murthy \& Masrur 2017~\cite{murthy2017subplatooning}\end{tabular} & Emergency Brake                                                                                                                          & Coordinated emergency braking in platoons considering the weakest vehicle in the platoon                                                                                      & R    \\ \hline
		Naranjo et al. 2008 /cite{naranjo2008lane}                                                                                                           & LaneChange                                                                                                                               & Applying fuzzy controller lateral and longitudinal control in LaneChange manoeuvre for overtaking                                                                             & R    \\ \hline
		Rai et al. 2015~\cite{rai2015real}                                                                                                                   & \begin{tabular}[c]{@{}l@{}}\begin{tabular}[c]{@{}l@{}}Merge, \\ \\ LaneChange,\\   Overtake\end{tabular}\end{tabular}                    & \begin{tabular}[c]{@{}l@{}}Concept of virtual leader to command synchronised lane changes, \\ \\ collision avoidance through potential-field controller\end{tabular}          & R    \\ \hline
		Sun et al. 2003~\cite{sun2003efficient}                                                                                                              & \begin{tabular}[c]{@{}l@{}}\begin{tabular}[c]{@{}l@{}}LaneChange,\\ \\   PlatoonChange, \\   GapOpen, GapClose\end{tabular}\end{tabular} & \begin{tabular}[c]{@{}l@{}}Manoeuvre to directly switch from one platoon to other platoon on \\ \\ adjacent lane, controller design for gap making and closing\end{tabular}   & R    \\ \hline
		Swaroop \& Yoon 1999~\cite{swaroop1999integrated}                                                                                                    & Emergency Brake and Emergency LaneChange                                                                                                 & \begin{tabular}[c]{@{}l@{}}Controller that couples braking and lane changing in emergency situations \\ \\ through V2V communication for higher safety\end{tabular}           & R    \\ \hline
		Usman \& Kunwar 2009~\cite{usman2009autonomous}                                                                                                      & Overtake                                                                                                                                 & Methodology for online generation of driving trajectories for overtaking manoeuvres                                                                                           & R    \\ \hline
		Wang et al. 2017~\cite{wang2017developing}                                                                                                           & JoinMiddle, GapOpen, GapClose                                                                                                            & Trajectories for JoinMiddle for minimizing carbon emissions                                                                                                                   & R    \\ \hline
		\multicolumn{4}{|l|}{* Focus of the paper, C = Communication, R = Regulation Layer, P = Platooning Layer}                                                                                                                                                                                                                                                                                                                                                                              \\ \hline
	\end{tabular}
}
\end{table*}

Another very important aspect of platooning is communication. 
The Join Tail-protocol description in \cref{fig:jointailcombined} shows the necessity for message transfers between participating vehicles. 
V2V standards, however, serve only collaborative awareness and are not sufficient for this case of collaborative driving. 
Researchers have developed various ways of formalising this communication while complying with the existing standards.
The authors of~\cite{bergenhem2015approaches} propose two protocols: A Minimal Protocol and a Full Platooning Protocol. 
While the Full Platooning Protocol is equipped for secured two-way communication, the Minimal Protocol is only suitable for one-way messages. 

The Minimal Protocol can be used for collaborative awareness as well as manual platooning, meaning that a driver must manually initiate the joining or leaving of a platoon. 
Within the platooning protocol, the authors differentiate among three types of messages: Service Announcements, Service Requests, and Control Data Messages. 
Service announcements are sent by platoon leaders to advertise services, for instance, the availability to add another vehicle to the platoon. 
Service requests are, for instance, sent by free vehicles to platoon leaders to express the willingness to join. 
Control Data Messages are sent periodically by platooning vehicles to maintain and update the integrity of the platoon.

In~\cite{amoozadeh2015platoon}, the WAVE Short Message Protocol (WSMP)~\cite{ieee16090-2013} is applied to carry information on the CCH. 
The messages are of two types. 
The first type are beacons. These beacons, like CAM, carry information about the vehicle state, e.g. the position, acceleration, and lane. 
Additionally, the beacon carries the platoon ID and the current vehicle position in the platoon if the vehicle is platooning. 
The position starts with 0 at the platoon leader and increments with each vehicle in the platoon counting upstream. 
The second type is micro-commands. 
These are used to initiate and control platoon manoeuvres. 
While beacons are usually broadcasted, most micro-command messages are unicasted or, if the manoeuvre involves multiple vehicles, multicasted. 
The authors provide a set of seventeen micro-commands to enable a multitude of manoeuvres.
To give the reader a more complete overview \cref{Table:platooning_literature} summarizes the research efforts involving studying platoon manoeuvres that were considered for this paper.

%
%
%
%
%
\section{Requirements and Principles}
\label{sec:requirements}
%

\begin{figure}[h]
	\centering

	\includegraphics[trim={0 0 0 0}, clip,width=0.45\textwidth]{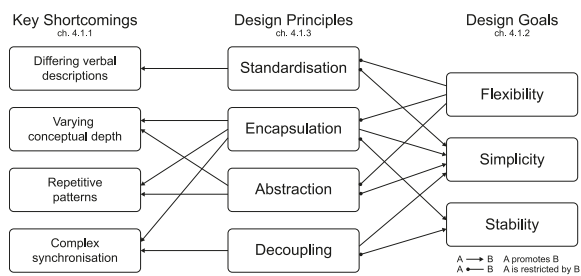}

	\caption{Relations of Key Deficits, Design Principles, and Design Goals for creating the SEAD framework.}
	\label{fig:principles}

\end{figure}

To relief the shortcomings described in the previous section, namely the challenges with comprehensibility, compilation and comparison of manoeuvre-descriptions, we propose a framework for the universal description of manoeuvres that builds on the four principles of Standardisation, Encapsulation, Abstraction, and Decoupling (SEAD). 
This framework will enable researchers and engineers to easily define new or alternative manoeuvres within the platooning layer \cref{fig:episode}.
First we will derive requirements for the framework design from the identified limitations, then we will define goals that the framework should fulfil, and postulate the overarching design principles to achieve the design goals. \cref{fig:principles} provides an overview of the interrelations among deficits, goals, and principles.

The SEAD design framework aims to resolve the major shortcomings that state-of-the-art manoeuvre descriptions face.  
The subsequent list summarises the identified deficits and defines the scope of problems the proposed framework should solve.
\begin{itemize}
	\item Varying conceptual depth:	Varying levels of conceptual depth within and among manoeuvre descriptions require to re-frame the manoeuvres before comparison.
	\item Differing verbal descriptions: Differing verbal descriptions for equivalent conceptual components require the manual alignment of terms before comparison.
	\item Repetitive patterns: Repetitive action-communication-patterns within and among manoeuvres make manoeuvres  more complex to read and tedious to be described.
	\item Complex synchronisation: The complex synchronisation between multiple role-specific FSM in one manoeuvre is challenging. This makes it hard to understand the collaborative aspect of a manoeuvre. Besides, it poses the threat of unstable states due to unforeseen circumstances.
\end{itemize}


For the SEAD framework to achieve practicality while resolving these deficits, we defined three design goals. 
The goals outline how we define practicality for the framework and give guidance in making all design decisions. 
The three qualitative goals are Flexibility, Simplicity, and Stability. 

\begin{itemize}

\item Flexibility: The framework must be capable of describing all manoeuvres in the current state of literature. Any design decision, thus, must ensure that the flexibility of describing manoeuvres is maintained and that restrictions are minimised.
\item Simplicity: The prime use case of the framework is reading and formulating new or alternative manoeuvres. Any design decision, thus, must promote the simplicity of reading and formulating manoeuvres.
\item Stability: The stability of manoeuvre designs is crucial, meaning that the design shall not allow for manoeuvres that end in an undefined state if unexpected driving situations occur. Any design decision, thus, must foster the stability of manoeuvres for any possible interruption or exception.
\end{itemize}


Given the postulated requirements, we derived four major design principles that help to resolve the current shortcomings and to fulfil the design goals. 
These can be summarised under the four terms Standardisation, Encapsulation, Abstraction, and Decoupling. 
The subsequent sections explain these principles in greater detail and interrelate them with the established design goals and identified deficits. 

\textbf{Standardisation}

As stated above, one prime issue in defining manoeuvres is the vastly varying terminology for equal conceptual components.
These components can be states within an FSM, V2V platooning messages or inherent reasons for state transitions such as autonomous decisions. 
Our framework resolves the lack of standardisation by introducing a finite set of symbols for these conceptual components. 
More specifically, the SEAD framework provides a syntax for describing vehicle actions, messages, and internal state transition triggers. 
The main requirement for this standardisation is the universal comprehensibility regardless of context or background.

The principle of Standardisation will mainly be affected by the design goals Flexibility and Simplicity. 
Although Standardisation will introduce syntactic rule sets, the standardised framework must be capable of expressing a sufficient conceptual depth not to obstruct the Flexibility goal. 
Besides, the Standardisation should also promote Simplicity, meaning that the standardised terms must be intuitively comprehensible and easy to grasp without the need for extensive preparation. 
The aim is that a person with no prior experience with the SEAD framework  must be able to understand the manoeuvres formulated using it.

\textbf{Encapsulation}

Throughout manoeuvre descriptions, we identified various repetitive patterns that include vehicle actions and inter-vehicle communication. 
The principle of Encapsulation will help pack these patterns into reusable blocks whereas one block may contain autonomous activities of one vehicle or synchronised actions of multiple vehicles. 
Moreover, these sub-manoeuvres will mainly comprise actions and states of equal conceptual depth. 
This helps in resolving the problems of varying conceptual depth, repetitive patterns, and complex synchronisation.

The Encapsulation will be influenced by all three design goals. 
Following the design goal of Flexibility, the Encapsulation of intertwined action-communication patterns must not impose manoeuvre-design restrictions. 
Thus, sub-manoeuvres on the lowest level must express the most fundamental patterns exhaustively. 
Thereby, Encapsulation will greatly promote Simplicity as it will allow reusing behavioural patterns without the need for redesigning them. 
This requires equipping sub-manoeuvres with a suitable interface that allows for integrating them into higher-order structures. 
Lastly, to achieve Stability, the sub-manoeuvres must be designed to always lead to predictable outcomes that can be handled by any context.

\begin{figure*}[!h]
	\centering
	\includegraphics[trim={0 0 0 0}, clip,width=0.75\textwidth]{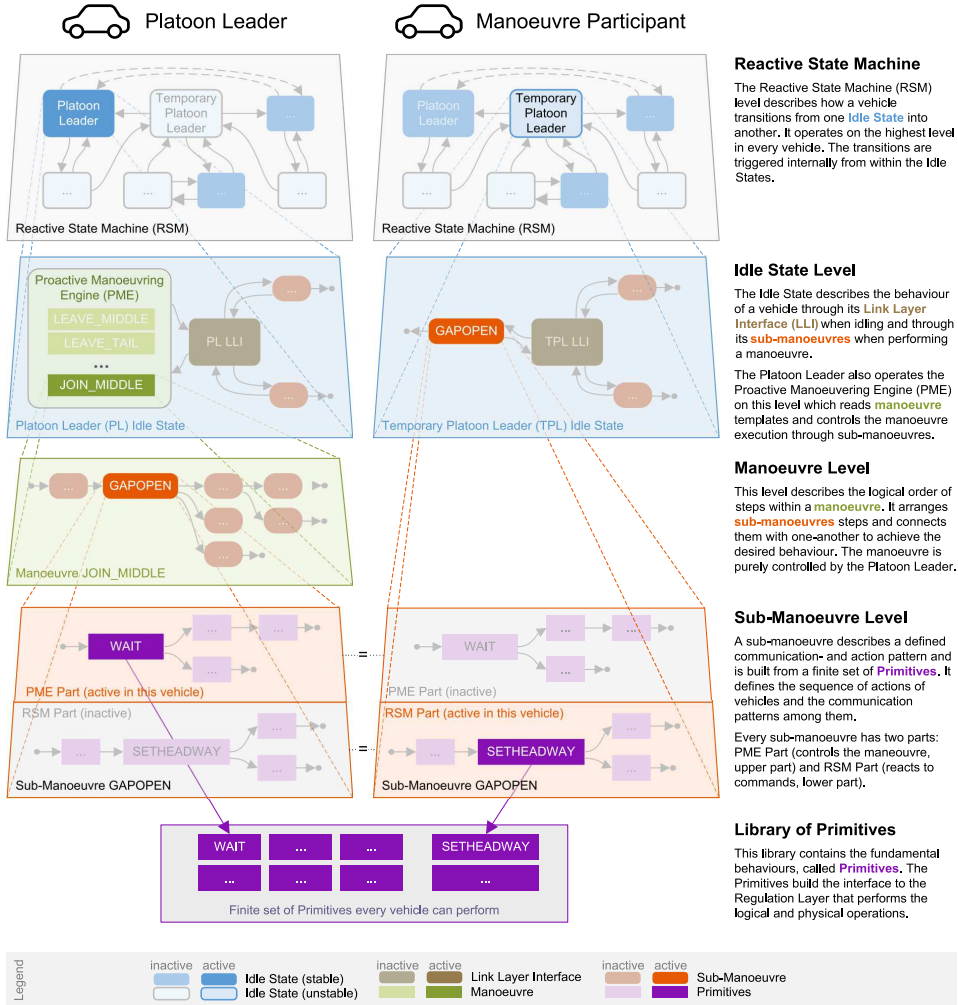}
	\caption{Overview of the hierarchical structure of the SEAD Framework. The figure provides an example where two vehicles, a platoon leader (left) and a temporary leader (right), perform a collaborative manoeuvre.}
	\label{fig:overview}
\end{figure*}

\textbf{Abstraction}

Extending Encapsulation, the principle of Abstraction will allow to reuse the encapsulated sub-manoeuvres within structures of conceptually higher levels and to encapsulate these higher order structures into reusable blocks. 
This recursive re-usage allows for arbitrarily complex structures whereas every level of conceptual depth can be designed separately without the need to operate on different levels at once. 
Abstraction helps to resolve the deficits of varying conceptual depth and handling repetitive patterns. 

As every system that re-frames complexity through hierarchical depth, our framework imposes certain restrictions.
Nevertheless, it shall fulfil the design goal of Flexibility, meaning that every abstraction will be carefully evaluated.
On the other hand, the Abstraction principle fosters Simplicity as the major part of the manoeuvre design process takes place at the higher levels of abstraction. 
This allows building highly complex manoeuvres with relatively low effort. 

\textbf{Decoupling}

Lastly, the Decoupling principle will allow describing complex manoeuvres entailing two or more vehicles from the perspective of the platoon leader. 
The SEAD framework, and specifically the Encapsulation principle, ensures that this single-sided description suffices to describe the behaviour of all vehicles. 
More specifically, the utilized master-slave structure leads to a general universal description of the reactive behaviour of the slave (platoon followers or free vehicles) while manoeuvres are only defined and steered by the master (platoon leader). 
This principle, in consequence, resolves the need for synchronising multiple manoeuvre descriptions.

Decoupling will considerably facilitate the process of reading and describing manoeuvres as both actions must only be performed from the leader perspective. 
This greatly drives the goal of Simplicity. 
However, designing the reactive behaviour of the slave-vehicles must be very comprehensive and elaborate in order to fulfil the design goal of Flexibility. 
%


\section{Framework Building Blocks}

In this section, we will give a detailed description of the building blocks that make up the SEAD framework.
Before we present each building block in a bottom-up fashion, we will describe a few properties of the the system to facilitate its overall comprehension:

\textbf{Scope:} The described platooning framework does not include high level decision-making processes such as when to form a platoon or when to leave a platoon.
This is the task of the Link Layer Interface (LLI) and could either be computed inside an autonomous vehicle or be given in the form of commands issued by another controlling entity. The framework at hand will only cover the platooning layer itself.

\textbf{Stability:} To ensure stability of the system, each platooning manoeuvre has to end in a stable state for all involved vehicles, regardless of manoeuvre's success.
We refer to these states as 'stable idle states' in contrast to 'unstable idle states' which are states during a manoeuvre when a vehicle is waiting for an action of another vehicle.
These stable idle states can be Platoon Leader (PL), Platoon Follower (PF), or Free Vehicle (FV).
When a vehicle is waiting for an instruction or action from another vehicle while performing a manoeuvre their respective unstable idle state would be WPL, WPF, and WFV.
Additionally, we make use of the unstable idle state Temporary Platoon Leader (TPL) to increase stability~\cite{segata2014supporting} when manoeuvres are aborted.
Only when a vehicle is in a stable idle state (i.e., not currently performing a manoeuvre) is it available to receive commands from the LLI.

\textbf{Control:} For every manoeuvre, we assume it to be carried out in a master-slave fashion, that is, one vehicle (naturally, the platoon leader) issues orders to the other vehicle(s). This does, however, not mean that manoeuvres can only be initiated by the platoon leader, it merely describes the control flow once the manoeuvre has started.

\textbf{Communication} We assume the existence of an underlying communication system that provides message primitives such as described by~\cite{amoozadeh2015platoon}. These messages include Requests (REQ), orders (ORD), done-confirmation (DN), abort (ABT), and acceptance/rejection (ACK/NACK). We abstract away from modelling the physical transmission of messages and assume perfect communication. To implement the concept of a Temporary Platoon Leader (TPL), an additional specialised message type TMPL SPLIT forces the split of a TPL and aborts the manoeuvre that is currently being processed.
REQ, ORD and DN messages can include additional information as required by the system, e.g., the size of the gap a vehicle is ordered to open.
We further assume that the underlying protocols can ensure that the successful transmission of messages.
If this cannot be ensured, then the sub-manoeuvres can be extended by adding respective time-out states and abort transitions.

\subsection{Framework Overview}

\cref{fig:overview} provides an overview of the SEAD framework and illustrates its hierarchical structure according to the paradigm of hierarchical state machines. 
The entire framework could be expressed as one big state machine, however, the transitions and states are designed in a way to enable all four SEAD principles (Standardisation, Encapsulation, Abstraction, and Decoupling).
The different layers can therefore be seen as different zoom levels or views of the platooning behaviour definition for autonomous vehicles.
In the following sections, we will introduce and explain the framework in a bottom-up fashion.

\subsection{Action Primitives}

The lowest layer of the framework is composed of action primitives which are actions that directly affect the physical or logical state of a vehicle.
We differentiate between physical primitives, state primitives, and other primitives.
The physical primitives are the direct interface to the regulation layer and affect the physical state of a vehicle, i.e, the vehicle's position or speed.
State primitives affect the role of the vehicle, i.e. whether it acts as a platoon leader or platoon follower.
They are needed to transition a vehicle from and to different idle states (e.g., at the end of a manoeuvre or when the vehicle needs to wait for instructions from another vehicle).
Lastly, other primitives are needed to either send messages, update platoon information, or wait for certain events.
Primitives should be designed in an orthogonal fashion, meaning that one primitive cannot be expressed by a combination of any other primitives. 
\cref{Table:primitives} provides an overview of action primitives that could be found in the state-of-the-art manoeuvre descriptions.

As the SEAD framework describes all actions on the platooning layer, the physical primitives provide directions to the low-level control of the vehicle instead of controlling the longitudinal dynamics directly. 
For instance, instead of setting the desired speed or acceleration of the vehicle, the primitives set the desired location. 
The path-following logic on the Regulation Layer will convert the target location into actionable commands (accelerate, decelerate, steer left / right) for the physical layer.

\begin{table*}[h]
	\caption{List of Action Primitives derived and condensed from the state of the art.}
	\label{Table:primitives}
	\resizebox{\textwidth}{!}{%
		\begin{tabular}{|l|l|l|l|l|}
			\hline
			\textbf{Acronym} & \textbf{Primitive} & \textbf{Parameters} & \textbf{Description} & \textbf{Actor}\\\hline
			\multicolumn{5}{|l|}{\textbf{Physical Primitives}}\\
			\hline
			MTP   & Move to position & TR: Target object,   RO: Relative offset & Move to a position defined by a relative offset to a target (vehicle) and align speed & FV, PL, TPL   \\ 
			\hline
			SH & Set headway & TH: Time headway \textit{or} SH: Space headway  & Set the desired time or space ahead to preceding vehicle and adjust real distance & PF, TPL       \\ 
			\hline
			\multicolumn{5}{|l|}{\textbf{State Primitives}} \\ \hline
			BFV & Become free vehicle & & Transitions the vehicle into a Free Vehicle & All but FV  \\ \hline
			BPL & Become platoon leader & & Transitions the vehicle into a Platoon Leader & All but PL   \\ \hline
			BPF & Become platoon follower & & Transitions vehicle into a Platoon Follower & All but PF    \\ \hline
			BTL & Become temporary leader  & & Transitions the vehicle into a Temporary Leader & All but TPL   \\ \hline
			SW & Set Wait & & Sets idling state to corresponding waiting state& FV, PL, PF\\ \hline
			USW & Unset Wait & &  Sets waiting state to corresponding idling state & WFV, WPL, WPF \\ \hline
			\multicolumn{5}{|l|}{\textbf{Other Primitives}} \\ \hline
			W  & Wait & E: Event to wait for, TO: Timeout & Waits for an event to occur, a message, or for a timeout & All \\ \hline
			SND & Send Message & M: Message, R: Receiver & Sends a message to the receiver & All           \\ \hline
			UPI & Update platoon information & & Updates information about the platoon (number and order of followers etc.) & PL\\ \hline
			\multicolumn{5}{l}{*(W)FV: (Waiting) Free vehicle, (W)PF: (Waiting) Platoon Follower, (W)PL: (Waiting) Platoon Leader, TPL: Temporary Platoon Leader} \\
		\end{tabular}%
	}
\end{table*}

\subsection{Formulating Sub-Manoeuvres}

With the list of idle states and action primitives, we can now create sub-manoeuvres.
A sub-manoeuvre encapsulates reusable behavioural patterns that involve two or more vehicles and transitions at least one of the participating vehicles from one idle state into another.
For each participating vehicle, the behaviour is described through a sequence of primitives which constitute a sub-state machine. 
Sub-manoeuvres must design action-communication patterns with the smallest reasonable scope to promote re-usability and therefore achieve the goal of Flexibility.

To implement the principle of Decoupling, each sub-manoeuvre is split into two or more sub-state machines, one for each vehicle participating.
As SEAD follows the master-slave paradigm, one sub-state machine controls the sub-manoeuvre (executed by the platoon leader), the other sub-state machines purely react.
The sub-state machines are connected and synchronised through V2V messages. 
This structure promotes the goal of Simplicity through the principles of Encapsulation and Abstraction: Any sub-manoeuvre can be reused without a thorough understanding of the inner workings once its behaviours and outcomes are defined.
We will discuss this separation and decoupling extensively in the Reactive State Machine (RSM) and the Proactive Manoeuvring Engine (PME) sections.
The controlling sub-manoeuvre is denoted with PME while the reactive ones are marked with RSM (as in \cref{fig:overview}).

To fulfil the goal of Stability, sub-manoeuvres must be designed such that any possible scenario (success or abort) leads to a defined outcome for every participating vehicle. 
Therefore, if one vehicle encounters a situation that will prevent the successful completion of the sub-manoeuvre and causes an abort-result, V2V communication (or time-outs) must cause all other sub-state machines to terminate at the same abort-result. 
Due to a shared understanding of the sub-manoeuvre among all vehicles, every vehicle is informed about the final state of all participants.
 


\begin{figure}[h]
	\centering
		\includegraphics[trim={0 0 0 0}, clip,width=0.5\textwidth]{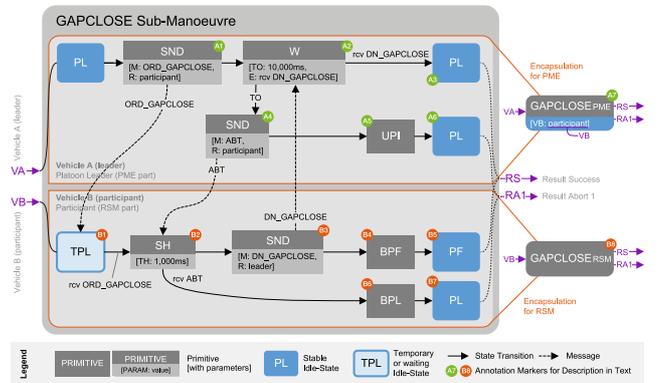}
		\caption{Sub-manoeuvre description and encapsulation for GAPCLOSE.}
	\label{fig:gap_close}
\end{figure}

\cref{fig:gap_close} illustrates an example sub-manoeuvre where a platoon leader orders another vehicle in the platoon to close the gap.
For stability reasons, the vehicle that closes the gap is a temporary platoon leader, so that when the sub-manoeuvre fails, it will be the platoon leader of all other vehicles behind.
The sub-manoeuvre includes two participants, the platoon leader (PL, vehicle A) and the temporary platoon leader (TPL, vehicle B).
The sub-manoeuvre can conclude in either Success (RS) or Abort 1 (RA1).

A initiates the sub-manoeuvre through commanding B to close the gap by sending an ORD GAPCLOSE message (A1) via V2V communication. 
After sending the message, A waits for the completion (A2). 
The message triggers B to execute the GAPCLOSE sub-manoeuvre (B1). 
B sets its headway to the requested intra-platoon distance and the regulation layer starts to decrease the distance until it reaches the desired headway (B2).
 
In the success scenario, the desired headway is reached, B signals the completion of closing the gap to the PL through sending a DN\_GAPCLOSE (B3) and transitions into the stable PF Idle State (B4). 
This concludes the sub-manoeuvre for B with a success-result RS (B5). 
A receives the message and concludes the sub-manoeuvre with a success-result RS (A3).

In the abort scenario, if closing the gap is taking too long, a timeout in A aborts the sub-manoeuvre. 
The timeout causes A to send an ABT message to B (A4) and to update the platoon information (A5). 
Afterwards, the sub-manoeuvres concludes for A with an abort-result RA1 (A6). 
B receives the message and will initiate the sequence to split from the original platoon by transitioning into a PL (B6). 
Once B is a PL, the sub-manoeuvre also concludes for B with an abort-result A1 (B7). 

The upper part (the pro-active part) and the lower part (the reactive part) of the sub-manoeuvre is encapsulated into two reusable blocks (A7 and B8, respectively).
These buildings blocks can then be used to build manoeuvres for the Proactive Manoeuvring Engine (PME) and the Reactive State Machine (RSM).
Due to the structure of the sub-manoeuvres, the initiation of a PME part will always leads to initiation of the RSM part as well. 
This principle allows defining manoeuvres from the perspective of the platoon leader without the need to describe the participant's behaviour.       
In the same fashion as in \cref{fig:gap_close}, it is possible to define a comprehensive set of sub-manoeuvres that allows assembling complex manoeuvres with limited number of restrictions.
We have created an online repository where we have made graphical depictions of manoeuvres and sub manoeuvres as well as their machine readable descriptions (see Section ~\cref{sec:language}) publicly available~\footnote{The library can be found at https://github.com/sead-framework/manoeuvre-catalogue}.

\subsection{Formulating Manoeuvres}

\begin{figure}[h]
	\centering
	\includegraphics[trim={0 0 0 0}, clip,width=0.5\textwidth]{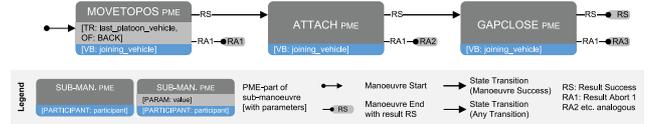}
	\caption{Description of the JOIN TAIL manoeuvre according to the PME logic.}
	\label{fig:join_tail}
\end{figure}

To promote the design goal of Simplicity and due to the master-slave paradigm of SEAD, a manoeuvre only needs to be described from the perspective of the platoon leader.
The behaviour of the other participants is defined in the sub-manoeuvres and the initiation of these sub-manoeuvres is done by the platoon leader via V2V messages.
A sequential manoeuvre (see \cref{sec:sim} how to define simultaneous manoeuvres) is a chain of sub-manoeuvres, where the transition to the next sub-manoeuvre depends on the result of the previous one.
\cref{fig:join_tail} shows the formulation of the JOIN TAIL manoeuvre as it was also described in \cref{fig:jointail}. 

In this illustration, the chain of sub-manoeuvres describes the entire manoeuvre. 
The blue lower part of the sub-manoeuvre boxes specifies the participating vehicles according to the notation in \cref{Table:primitives} whereas vehicle A (VA) is not specified since it is always part of the manoeuvre as the PL, and vehicle B (VB) is the participating vehicle. 
The JOIN TAIL manoeuvre requires no additional actions in case of an abort and will conclude in a stable state achieved through the abort-architecture within the sub-manoeuvres.
More elaborate manoeuvres such as the JOIN MIDDLE require more sophisticated abort structures (e.g., a GAP CLOSE for a platoon follower if the joining vehicle was unable to lane change into the platoon).
This is necessary to ensure that no vehicle will be in an unstable idle state once the manoeuvre is finished.

Defining manoeuvres in this fashion largely promotes Simplicity as a manoeuvre can be described from the leader's point of view while the reactive part of all sub manoeuvres (also referred to as the universal RSM) takes care of the  participating vehicles' perspective.

\subsubsection{Simultaneous Manoeuvres}
\label{sec:sim}
Manoeuvres with two or more participating vehicles can potentially benefit from the parallel execution of sub-manoeuvres.
To facilitate this, the SEAD framework introduces a wrapper for the simultaneous execution, referred to as the SIM WRAPPER. 
This construct can involve an arbitrary number of simultaneous sub-manoeuvres whereas every sub-manoeuvres’ controlling part (the upper part in~\cref{fig:gap_close}) is executed by the same leader and the reactive portion (the lower part in the same figure) is executed by the participants. 
Two simultaneous sub-manoeuvres, however, cannot involve the same participating vehicle. 

To comply with the requirement that a state machine can only be in one state, the SIM WRAPPER can be understood as a product state machine of the controlling portions of all involved sub-manoeuvres. 
Since the reactive portion is executed in separate vehicles through the Decoupling principle, they occur in separated state machines in distinct systems. 
The execution result can be any element from the Cartesian product of all execution result sets from all sub-manoeuvres. 

 
\subsection{Idle States and Super-States}

\begin{figure}[h]
	\centering
	\includegraphics[trim={0 0 0 0}, clip,width=0.5\textwidth]{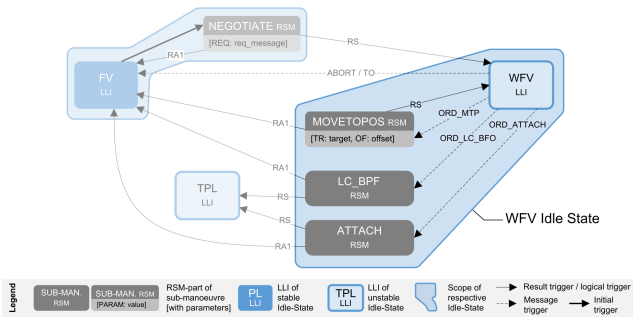}
	\caption{Definition of the WFV Idle Superstate. It contains the WFV LLI Idle state  as well as the sub-manoeuvres that can be performed by the WFV to transition into another Idle State. }
	\label{fig:wfv}
\end{figure}

With the definition of idle states and sub-manoeuvres, it is possible to define them together to derive a state-machine on an abstraction level that clearly shows how the vehicle can transition from one idle state to another via which sub-manoeuvre. 
We combine the idle state and its associated sub-manoeuvres (i.e., the sub-manoeuvres that a vehicle can execute if it is in the idle state upon reception of a V2V message) into an idle super-state.
This concept is shown in \cref{fig:wfv}, where the idle state WFV (Waiting Free Vehicle) and three sub-manoeuvres LC\_BPF (Lane Change \& Become Platoon Follower), MOVETOPOS (Move to Position), and ATTACH are all combined into a superstate.
This superstate can only be left through the successful or unsuccessful execution of sub-manoeuvres or when a time-out occurs.

Every stable Idle State (FV, PF, PL) has an idling superstate.
Within this superstate, the idle state will be referred to as a LLI (Link Layer Interface) idle state, because in these states, the vehicle can make (or receive) high-level decisions to carry out collaborative actions, for instance, joining or leaving a platoon, the decisions for which are made in the Link Layer. 
When the vehicle is in an unstable idle state (i.e., WFV, WPF, WPL, TPL), thus in a manoeuvre, the LLI is inactive because manoeuvre initiation is only possible when the vehicle is not already performing a manoeuvre. 
Since this paper focuses on the Platoon Layer and the LLI models the Link Layer, the inner workings of the LLI will not be covered here. 


\subsection{Reactive State Machine (RSM)}

\begin{figure*}[t]
	\centering
	\includegraphics[trim={0 0 0 0}, clip,width=0.8\textwidth]{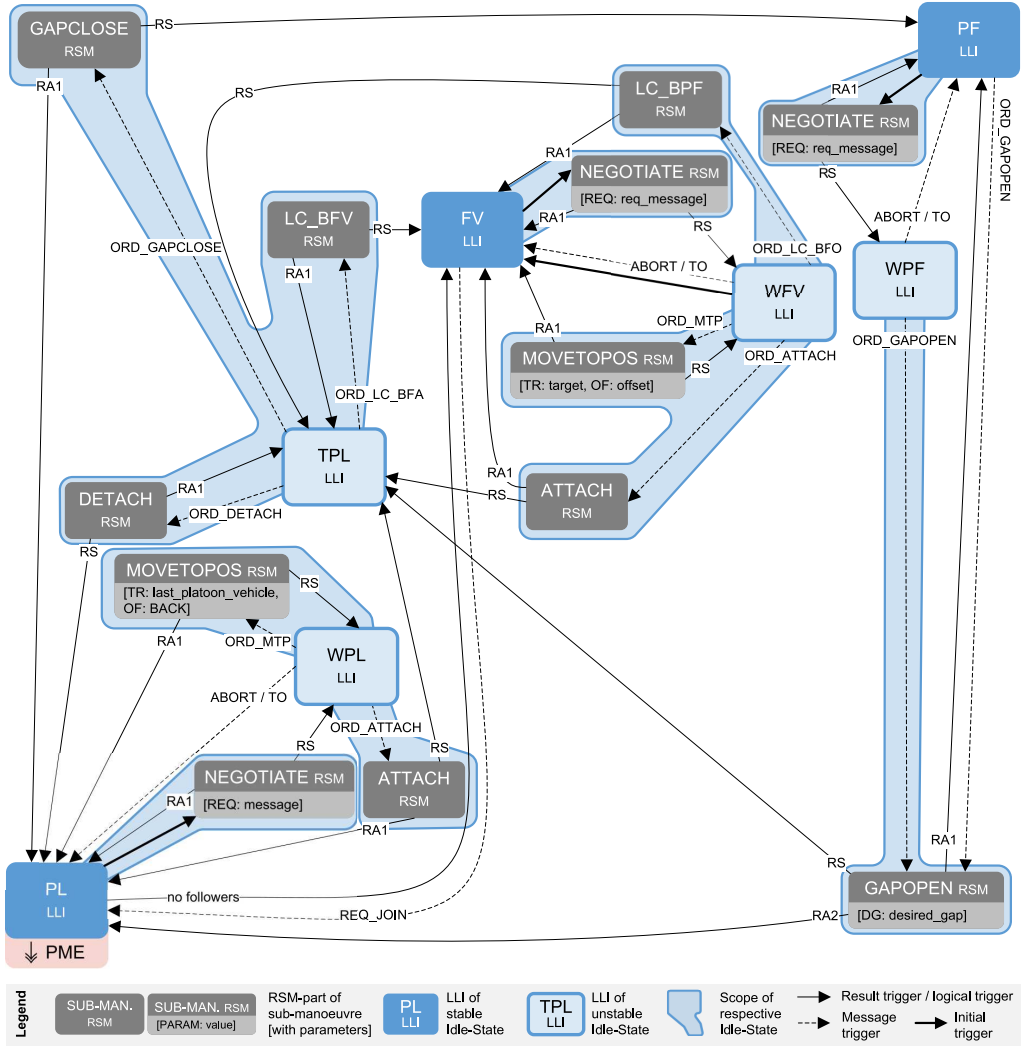}
	\caption{Reactive State Machine (RSM). It describes the reactive behaviour by combining all sub-manoeuvres.}
	\label{fig:rsm}
\end{figure*}

The combination of all idle superstates into one big interconnected state-machine yields the so-called Reactive State Machine (RSM), which can be seen as a complete behaviour definition for all platooning vehicles except the platoon leader.
As the platoon leader coordinates and controls the manoeuvres, this state-machine is called \textit{reactive} as it reacts to what the leader is doing.
Figure~\cref{fig:rsm} shows the complete RSM for a platooning system which supports a number of basic sub manoeuvres.
For better readability, we chose sub manoeuvres as the abstraction level in this illustration, however, each of the sub manoeuvre boxes could be replaced by the entire sub manoeuvre definition (e.g., \cref{fig:gap_close}.

The RSM describes the universal reactive behaviour of any vehicle for any manoeuvre that can be built using sub-manoeuvres as building blocks. 
However, the RSM describes only the decoupled reactive behaviour (as indicated by the RSM in any sub-manoeuvre box).
It does not implement how the manoeuvre is logically performed, i.e. the sequence of sub-manoeuvres comprising a manoeuvre. 
This implements the first half of the Decoupling principle. 
The second half of Decoupling, the formulation and control of manoeuvres, is implemented by a complementary structure that steers manoeuvres, namely the Proactive Manoeuvring Engine (PME).

The introduction of the RSM strongly promotes the goal of Stability. 
As can be seen in \cref{fig:rsm}, the RSM defines the behaviour in case of an abort for every sub-manoeuvre. 
This structure, thus, always brings the vehicle to a defined state in whichever way a manoeuvre terminates. 
According to the Abstraction principle, the RSM reuses sub-manoeuvres, which leads to all building blocks in the RSM being on an equivalent conceptual level. 
This directly mitigates the two key shortcomings of varying conceptual depth and repetitive patterns. 
Since the structure and design of the sub-manoeuvres is aligned with the goal of Flexibility, the RSM introduces no further restrictions regarding this concern. 

\subsection{Proactive Manoeuvring Engine (PME)}

While both proactive and reactive behaviour definitions can be combined into one state-machine, we separate them for the sake of Simplicity and Decoupling.
To this end, the Proactive Manoeuvring Engine (PME) complements the reactive behaviour of the RSM (as can be seen in the left bottom of \cref{fig:rsm} they are indeed connected).
The PME is therefore an extension to the PL’s RSM behaviour and is responsible for the coordination of platooning manoeuvres. 

\begin{figure*}[h]
	\centering
	\includegraphics[trim={0 0 0 0}, clip,width=0.8\textwidth]{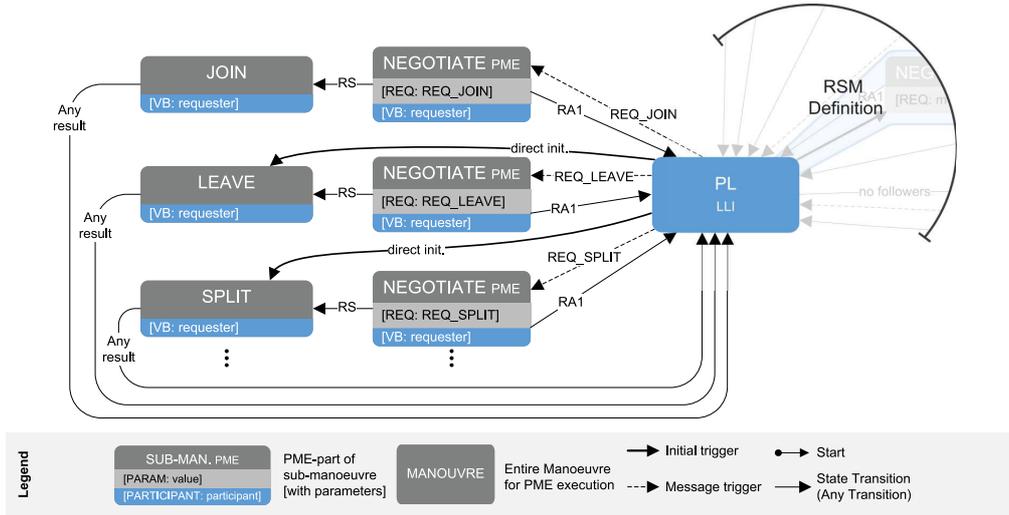}
	\caption{Proactive Manoeuvring Engine (PME). A manoeuvre can either be initiated through an acknowledged request through NEGOTIATE or directly through the LLI of the PL.}
	\label{fig:pme}
\end{figure*}

\cref{fig:pme} shows the PL LLI with the connection to the RSM part (\cref{fig:rsm}) on the right side and the entire PME on the left side.
The level of abstraction chosen in this figure is on the manoeuvre layer with the example at hand supporting the JOIN, LEAVE and SPLIT manoeuvres.
These manoeuvre boxes could be extended to their respective contained sub-manoeuvres (or even further to include the entire controlling part of each sub-manoeuvre), however, Abstraction and Decoupling allow us to illustrate the platooning system in a more comprehensible way.
The PME combines the manoeuvre schemes of all manoeuvres and steers the manoeuvres from the perspective of the PL. 
Manoeuvres are initiated through the LLI of the PL either directly (direct init. in \cref{fig:pme}) or through a request REQ that triggers a NEGOTIATE sub-manoeuvre. 
When adding a new manoeuvre, rules in the LLI must define when it will be called (i.e. which REQ message evokes a NEGOTIATE or which conditions triggers a direct init.) as illustrated in \cref{fig:pme}.

Only manoeuvres that concern exclusively platoon followers can be instantiated directly by the PL. 
For instance, a PL cannot command an FV to join its platoon without a prior request of the FV to join. 
This idea imposes that joining is always initiated through an FV. 
However, by designing an additional sub-manoeuvre where the PL requests an FV to join, the SEAD framework can also adapt to this paradigm.

\subsection{Manoeuvre Design Language}
\label{sec:language}
Although the visual description of manoeuvres and sub-manoeuvres is easily comprehensible for humans, machines will not be able to process it. 
To allow flexibly redesigning manoeuvres and sub-manoeuvres, we have developed a Manoeuvre Design Language (MDL) that directly translates from and into a graphical representation.
In the future, a graphical editor to create and export manoeuvres and sub-manoeuvres as JSON MDL files will help to easily design manoeuvres graphically and to directly feed them into simulation systems. 
The simulation system parses the MDL file and generates the code required for the execution of manoeuvres according to the SEAD framework. 
The MDL is based on JSON and was inspired by the syntax and structure of the Amazon States Language\cite{}. 
One JSON MDL file has a unique ID (or action ID) and describes a (sub-)manoeuvre following a fixed syntax.
The syntax of the language is outside of the scope of this paper but can be accessed in~\cite{deboeser2019towards}

\section{Conclusion and Future Work}
\label{sec:conclusion}
This paper introduces the SEAD framework that simplifies the formulation of manoeuvres for vehicle platooning. 
It is based on the four principles of Standardisation, Encapsulation, Abstraction and Decoupling (SEAD). 
As previous research has shown~\cite{varaiya1993smart}, vehicle platooning has the potential to substantially increase road capacity and traffic throughput, providing a potential solution for traffic systems to adapt to the ever-increasing traffic demand. 
Although vehicle platooning is a promising concept, it remains challenging to define and describe collaborative manoeuvres. 
This poses a bottleneck in the development pace of the platooning concept and hampers its applicability in real-world scenarios. 

The design of the SEAD framework was conceptualised to resolve four key shortcomings of the state-of-the-art description of manoeuvres using coupled state machines: First, as a manoeuvre description consists of one state machine per participant, the reader must synchronise the state machines to understand the manoeuvre. 
Second, the investigated schemas reveal varying conceptual depth within and among manoeuvre description as well as, third, differing verbal descriptions of equivalent components. 
This heterogeneity makes it difficult to understand and compare manoeuvres. 
Finally, various action-communication patterns surfaced in multiple manoeuvre descriptions, making them seem more complex than they are. 


The SEAD framework is a first step to formalise the Platoon Layer of an Automated Highway System. 
It provides the means to conduct further research on the performance of manoeuvre variations, alternatives and platoon forming strategies. 
There are two main future research fields involving the proposed framework.

%

\textbf{Improving and Extending the SEAD Framework}

Once the required simulation tools for platooning in urban environments are in place, further building blocks may be designed extending the SEAD framework. 
The framework will allow formulating complex urban manoeuvres such as LEAVE MIDDLE AND TURN LEFT (or LMTL), yet the further development of primitives, sub-manoeuvres, and communication patterns will be required. 

To increase the adoptability and usability of the SEAD framework, future research could focus on developing a stand-alone application-agnostic tool that models the Platoon Layer according to the presented framework. 
However, since the Platoon Layer and the Regulation Layer are closely interconnected through the Action Primitives, an elaborate communication interface between these two layers needs to be developed to allow for modularisation. 
The resulting module could then be optimised in simulations and, in the further future, be deployed within the vehicle software.

Furthermore, although the proposed framework allows defining elaborate abort structures for manoeuvres and sub-manoeuvres, it does not propose an abort-and-retry structure. 
Once elaborate models are in place to evaluate if a second attempt could be successful, higher-order manoeuvres and re-modularisation of certain sequential parts of a manoeuvre provide the opportunity to implement such abort-and-retry structures.

\textbf{Empowering Further Studies}

As mentioned before, the biggest advantage of the SEAD framework is its capability of designing manoeuvre variants and alternatives through re-arranging the sub-manoeuvres once all building blocks are implemented and the RSM defined all required state transitions.
This allows for various dynamic investigations.

For instance, traffic simulators implementing the framework such as BEHAVE~\cite{Ivanchev2019itswc} could provide a means to identify the most efficient alternative of a manoeuvre through simulating all possible alternatives and measuring the execution time, success rate, and traffic flow influence of the alternatives.
Using the same approach, different platoon formation strategies (Weakest-in-Front, Last-in-at-Tail, dynamic contextual strategies etc.) could be investigated regarding their influence on the overall traffic flow.
The SEAD architecture has already been implemented and is a critical part of the Autonomous Vehicle Driver Model architecture described in \cite{braud2021avdm}.

The proposed framework utilises a timeout-strategy so that manoeuvres are not leading into a deadlock if, for instance, a human-driven vehicle is blocking the way to complete a manoeuvre. 
The timeout-strategy, however, is a mechanism to ensure deadlock-freeness. 
Research may investigate further models to identify blocking situations with a low likelihood of manoeuvre success to interrupt the manoeuvre using as a trigger an event rather than a time-out.
Otherwise, simulations may allow to numerically optimise the time-out parameters with, for instance, the overall traffic flow as the objective variable to maximise.

After all, platoon manoeuvring is a powerful yet complex technique to fulfil the ever-increasing traffic demand with the given capacities we have. 
Further research will have to investigate many more gaps and find the most effective strategies to maximise traffic throughput. 
To unlock the full potential of platooning, the SEAD framework aims to pave the way for this future research by providing a formalisation of the platooning logic and simplifying the way how new manoeuvres are created and existing ones are compared and optimised.

\bibliographystyle{IEEEtran}
\bibliography{references}

\end{document}